Animated 3D Human Models for Use in Person Recognition Experiments

Jean M. Vettel[1,2,3], Justin Kantner[1,2], Matthew Jaswa[4], Michael Miller[2]

[1]U.S. Army Research Laboratory, [2]University of California, Santa Barbara, [3]University of Pennsylvania, [4]DCS Corporation

Corresponding Author:

Jean M Vettel

U.S. Army Research Laboratory

459 Mulberry Point Road

Aberdeen Proving Ground, MD 21005

410.278.7431

jean.m.vettel.civ@mail.mil

Abstract (max 250 words)

The development of increasingly realistic experimental stimuli and task environments is important for understanding behavior outside the laboratory. We report a process for generating 3D human model stimuli that combines commonly used graphics software and enables the flexible generation of animated human models while providing parametric control over individualized identity features. Our approach creates novel head models using FaceGen Modeller, attaches them to commercially-purchased 3D avatar bodies in 3D Studio Max, and generates Cal3D human models that are compatible with many virtual 3D environments. Stimuli produced by this method can be embedded as animated 3D avatars in interactive simulations or presented as 2D images embedded in scenes for use in traditional laboratory experiments. The inherent flexibility in this method makes the stimuli applicable to a broad range of basic and applied research questions in the domain of person perception. We describe the steps of the stimulus generation process, provide an example of their use in a recognition memory paradigm, and highlight the adaptability of the method for related avenues of research.

**Introduction**

Experimental research across numerous domains has employed 2D pictures of cropped human face stimuli to examine behavioral phenomena and/or neural correlates of performance on person recognition tasks. Many laboratories have developed and released standardized sets of face stimuli, including the MPI database (Troje & Bülthoff, 1996), the FERET database (Phillips, Moon, Rizvi, & Rauss, 2000), the AR database (Martinez & Benavente, 2000), the Radboud database (Langner et al., 2010), and the FEI database (Thomaz & Giraldi, 2010). Research has successfully utilized these stimulus sets across a diverse array of task paradigms, including our own work on criterion shifting in recognition memory (Aminoff et al., 2012), development of perceptual representations (Tanaka, Meixner, & Kantner, 2011), and neural features for face detection and individuation (Nestor, Vettel, & Tarr, 2012). Recently, there has been increased interest in connecting understanding of isolated face recognition with person identification processes more generally (Berlucchi, 2011; Campanella & Belin, 2007; Chan & Baker, 2011; Longmore & Tree, 2013; Minnebusch & Daum, 2009; Moro et al., 2012; Ramsey, van Schie, & Cross, 2011; Robbins & Coltheart, 2012; Stekelenburg & de Gelder, 2004; Yovel & Belin, 2013). In addition to identity judgments, researchers have highlighted the importance of face and body information for emotion identification (Aviezer, Trope, & Todorov, 2012b; Grèzes, Pichon, & de Gelder, 2007; van de Riet, Grezes, & de Gelder, 2009) and scene perception (Bindemann, Scheepers, Ferguson, & Burton, 2010; Righart & de Gelder, 2008). These findings reveal a tight interaction between face and body information for task performance, indicating the value of full body human stimuli for understanding person perception as it occurs in naturalistic environments.

Moving to stimulus sets of full-bodied individuals has inherent challenges. Taking pictures to generate large sets of 2D images is time consuming, and it is challenging to maintain consistent lighting, viewpoint, and emotional expression across individuals. Studies that have incorporated images of full human bodies have often used a limited number of unique items (Bindemann et al., 2010; Stekelenburg & de Gelder, 2004). Experiments examining the influence of scene context on person perception are challenging to implement from libraries of human photographs. Furthermore, the 2D stimulus approach precludes entire classes of experimental questions such as time-evolving emotional expressions, action perception, or social gestures and interaction.

One alternative to using photographs for person perception research is to utilize 3D human models. Research has validated the use of synthetic human stimuli in face perception research, demonstrating that such stimuli yield benchmark phenomena in face processing (Matheson & McMullen, 2011), action perception (Casile et al., 2009), and emotion perception (McDonnell, Jörg, McHugh, Newell, & O'Sullivan, 2009). However, commercially-available sets of 3D models typically contain a dozen or so models, and combining models from multiple sets is typically infeasible due to variations in model rendering or resolution, model clothing, and model pose across sets. Yet many experimental paradigms, including recognition and decision making tasks, require a hundred or more test trials with unique stimuli to obtain stable behavioral indices of performance. Even available procedural content generation approaches that can efficiently generate large numbers of stimuli (Hendrikx, Meijer, Van Der Velden, & Iosup, 2013) do not produce the type of variability in facial features that is desirable in many person perception experiments. Consequently, we developed a method to produce animated human models with individualized features for use in commercially-available 3D simulation environments.

Our approach unites the parametric control of facial features using FaceGen software with an industry-standard Cal3D format for human models that is compatible with many virtual 3D environments. We attached novel FaceGen head models to commercially-purchased 3D avatar bodies using the popular 3ds Max graphics program. Thus, our method capitalizes on software already common in cognitive research. We developed this process for use in an old-new recognition memory paradigm in which background scenery indicates the probability that a presented model is a studied human target (“old”), and we describe the generation of this stimulus set as an example application of the stimulus generation process. However, this stimulus generation framework is versatile and adaptable to numerous questions in person perception research.

## Method

Our stimulus generation process utilizes two commercial software packages: FaceGen Modeller (FaceGen) from Singular Inversions (Toronto, Ontario) and 3D Studio Max (3ds Max) from Autodesk (San Rafael, California). The process uses these software packages in conjunction with a purchased set of 3D human models, a purchased 3D simulation environment, and a collection of scripts and custom software to automate parts of the process. This section first describes the three main steps to generate unique 3D human models shown in Figure 1: creating unique head models in FaceGen, attaching the heads to purchased 3D bodies in 3ds Max, and generating models in Cal3D format. The section concludes with a brief discussion of application of this process for a recognition memory paradigm where 324 human models were embedded into a simulated 3D desert metro environment and exported as 2D images.

[Insert Figure 1]

*Step 1 in FaceGen: Synthetic Head Models*

Synthetic FaceGen head models have been used in face processing research for at least a decade (Shimojo, Simion, Shimojo, & Scheier, 2003). The package is built on the statistics of feature variations in human faces, including parameter ranges that differentiate males and females as well as parameter variations tied to age, expression, and ethnicity. The software has over 75 parameters to control the shape and relational distances of facial features and over 30 parameters to control the color and shading of the face. Thus, FaceGen stimuli have been used to study viewpoint (Chen, Yang, Wang, & Fang, 2010; Fang, Ijichi, & He, 2007), gender and race discrimination (Corneille, Hugenberg, & Potter, 2007; Matheson & McMullen, 2011; Papesh & Goldinger, 2010; Yang, Shen, Chen, & Fang, 2011), attractiveness (Shimojo et al., 2003; Xu et al., 2012), trustworthiness (Oosterhof & Todorov, 2008; Todorov, Baron, & Oosterhof, 2008; Verosky & Todorov, 2010; Xu et al., 2012), and recognition (Arcurio, Gold, & James, 2012; Mur, Ruff, Bodurka, Bandettini, & Kriegeskorte, 2010; Russell, Sinha, Biederman, & Nederhouser, 2006). FaceGen stimuli have also been used to examine face processing effects in clinical populations (Martens, Hasinski, Andridge, & Cunningham, 2012). Although most studies have used hairless head models, FaceGen does include several sets of hairstyles that can be applied to each of the generated heads.

FaceGen head models can be generated at random or they can be based on imported profile photos to generate personalized 3D models. This feature has been used to study familiarity (Verosky & Todorov, 2010), and future research could utilize this feature to create 3D avatars of known individuals. Building models personally familiar to the participant may increase an individual's immersive experience and facilitate more realistic behavior in simulated environments. An interesting use of this approach was in a popular online soccer league, where FaceGen was used to create avatars of current soccer players and incorporate realistic emotional expressions for the avatars (Visser, 2011). In our test of this software feature, models were easily generated from photos of faces personally familiar to us and then embedded in a desert metro environment as shown in Figure 2.

A critical feature of the FaceGen software for our method is the ability to export FaceGen head models in 3ds Max format. The export feature includes parameters for the head model's scale, xyz rotation, and xyz translation, and setting these parameters with values that are specific for the intended 3ds Max body streamlines the process to connect the exported head model with an avatar body.

[Insert Figure 2]

*Step 2 in 3ds Max: Manipulating 3D Human Models*

Many commercially-available 3D human models are designed with 3ds Max, and the native 3ds Max files are available for purchase and download from the host websites. Researchers can design their own bodies in 3ds Max (or a similar software program), but our process edits purchased 3D human models.

Our method first removes the head model from the purchased body model. For most purchased human models in our experience, the mesh skeleton for the head is a separate object from the mesh skeleton for the body, so it is easy to select the head and simply delete it. For any model with a joint head and body mesh, 3ds Max provides cutting tools to separate the head mesh and editing tools to reconstruct the shoulders and make a closed body mesh. The exported FaceGen head model is then imported into the 3ds Max scene and attached to model's underlying skeleton. This step ensures that the head will move properly when any of the model's animations are played since the animations are connected to the skeleton, not the mesh.

Another option to increase variability among the human models is to edit the clothing textures. Each model in our purchased set included a bitmap file that contained clothing textures that are wrapped around the 3D body skeleton. Using any image editing program (e.g., Photoshop, GIMP, etc), a color mask can be applied to the existing clothing texture, creating different clothing colors for each model. Researchers can also design their own clothing patterns or use procedural content generation methods to manipulate other aspects of the clothing texture such as 3-dimensional smoothness, transparency, and glossiness (Rhoades, Turk, Bell, Neumann, & Varshney, 1992).

*Step 3 in Cal3D: Creating Animated 3D Models*

Once the FaceGen head is attached to a body skeleton in 3ds Max, the new human model is exported as a Cal3D mesh. At this stage, the 3ds Max human model is without textures to color the skin, clothing, etc, so we wrote custom software to unite all of the Cal3D meshes, textures, and animations into a 3D model. The Cal3D format is compatible with simulation tools built on the popular Delta3D library, and there are many freeware tools to convert Cal3D into other popular 3D model formats for use in many standard 3D interactive environments. This step of the stimulus generation process could incorporate additional animations to customize the gestures and movement patterns for the models. Previous research by Casile and colleagues (2009) used data from motion-capture to animate the movement of an avatar, an approach applicable here that could benefit from the availability of several point light action databases (Manera, Schouten, Becchio, Bara, & Verfaillie, 2010; Zaini, Fawcett, White, & Newman, 2013) to study social gestures and/or action perception.

Tools also exist to embed the models in 3D environments, create particular scenes, and export 2D images for use in traditional laboratory experiments, a step we implemented in an example described below.

*Example Application: Recognition Memory of Human Targets*

We used this stimulus generation method to produce a 324-item stimulus set for a recognition memory paradigm. Our stimuli were developed from a purchased set of 3D human models ("Arabic Civilians" by ES3DStudios) and a purchased desert metro 3D simulation environment ("Arab Streets Stage-03" by ES3DStudios).

In our experiment, participants studied a subset of the models on white backgrounds and then were tested for their memory of the models shown embedded in a desert metro environment. Models stood in one of eight locations in the city or in one of eight locations in the outskirts of the city. The background context (city or outskirts) indicated the probability that the individual was studied ("old"), and the experiment examined how well study participants could incorporate this probability information into their recognition judgments. In order to encourage participants to decrease reliance on memory evidence in making recognition judgments, our stimulus set was designed such that each item contained a unique face but was highly confusable with many other items in the set. We validated the difficulty in discriminating “old” (previously presented) items from “new” items within the set by computing recognition sensitivity (*d’*) across 389 participants. We report those results after describing the process of developing the set.

The head models in our stimulus set fall within the parameters defined in FaceGen for Caucasian faces, where the settings in the Race Morphing continuum had a large European component and little or no African, East Asian, or South Asian component. The head models range between the ages of 20 and 40 based on FaceGen parameters for an Age continuum. The head models were designed to vary along two feature dimensions: sex and skin color. Sex is determined by the position of the color and shape sliders along a FaceGen Gender continuum ranging from “Very Female” to “Very Male.” Extreme points on the Gender continuum were avoided, as they tended to produce unnatural-looking faces. Skin color separability was achieved by constraining the Skin Shade setting (in the Colour tab) to a range of 0.8 to 1.2 for each light-skinned face and to a range of -0.8 to -1.2 for each dark-skinned face. Thus, each Caucasian head model, aged 20-40, falls into one of four categories: light-skinned male, dark-skinned male, light-skinned female, and dark-skinned female. The set contains 81 exemplars of each category.

Four of the hairstyles in FaceGen were selected for males and four for the females: midlength straight male, base short, short black, and preppy blonde for the males, and midlength straight, midlength messy, long curvy, and roman for the females. We used the available color variations for the hairstyles; thus, our stimulus set contains 17 different colors across female hairstyles and 7 different colors across male hairstyles.

Four male and four female bodies were selected from our set of purchased animated 3D models. We increased the variability in the bodies of the models by adjusting the color of the clothing. We created thirteen clothing colors for each of the models.

[Insert Figure 3]

We created a unique head model for each of our 324 stimuli and systematically varied the model body, the hairstyle and hair color, the skin color, and the clothing color across the set. A sample of variability across models is shown in Figure 3. Each of these 3D models was embedded in the purchased desert metro simulation environment, and a 400x600 pixel 2D image was produced of each model with three backgrounds. Models on a white background were used in the study phase of the recognition memory experiment, and in the test phase, models were shown standing in either a city location or a city outskirts location. Samples of embedded models are shown in Figure 4. Although not used in the current recognition memory paradigm, each Cal3D model is associated with several animated behaviors that can be used in interactive simulations. A subset of animations is shown in Figure 5.

[Insert Figure 4]

In summary, our stimulus set design emphasized three features: (1) an increase in stimulus complexity with full-bodied individuals, (2) an increase in task complexity with environmental scenes conveying probability information, and (3) an increase in the difficulty of the memory judgment with a relatively homogenous stimulus set. We evaluated the third parameter by computing recognition sensitivity (*d'*) across 389 participants. The *d'* measure was calculated as z(H) – z(FA), where H and FA are the hit and false alarm rates, respectively. As intended, old-new discrimination was very poor. Across all participants, the mean *d'* was 0.140 (corresponding to a mean hit rate of .54 and a mean FA rate of .48). This low *d'* was a critical feature to examine whether participants increase their reliance on probability information (as opposed to information in memory) when discrimination of old and new items is nearly impossible (Kantner, Vettel, & Miller, under review).

[Insert Figure 5]

**Conclusion**

The stimulus generation method described is a versatile tool for developing large stimulus sets of 3D human models for person perception research with parametric control over identity features and animated behaviors. The method capitalizes on commonly used software packages in perceptual research to elucidate substrates of behavior in increasingly realistic task paradigms. The approach is adaptable and allows careful tuning and realistic appearance of synthetic human models for use in virtual reality (Schultheis, Himelstein, & Rizzo, 2002; Tarr & Warren, 2002), simulated task (Gray, 2002) and virtual gaming (Visser, 2011) environments, and more traditional experimental paradigms (Matheson & McMullen, 2011).

The stimulus generation method is composed of three primary steps, and each can be adapted to generate animated 3D human models for many avenues of person recognition research. First, the parametric control of facial features in FaceGen enables research spanning diverse domains, including race perception (Papesh & Goldinger, 2010), gender discrimination (Yang et al., 2011), trustworthiness (Todorov et al., 2008), familiarity (Verosky & Todorov, 2010), and emotional expression (Cade, Olney, Hays, & Lovel, 2011), including the recently developed FACSGen toolbox (Roesch et al., 2011). Second, the editing capabilities in 3ds Max allow head models and body models to be interchanged and adapted for particular poses, animations, or appearance. The use of animated 3D human models holds promise for better understanding the interaction between face and body information for emotion identification (Aviezer, Trope, & Todorov, 2012a; Aviezer et al., 2012b), scene perception (Bindemann et al., 2010), and action perception (Casile et al., 2009). Finally, the versatility of the Cal3D format for incorporating human models into simulation environments and virtual worlds facilitates a host of interactive experiences to study more naturalistic behavior (Bainbridge, 2007; Ingram & Wolpert, 2011; Mathiak & Weber, 2006; Schultheis et al., 2002; Wilson, 2002).

More generally, our method enables increased stimulus and task complexity in person perception research. Tools for naturalistic experimental paradigms are essential for research that examines how well robust laboratory findings translate into increasingly complex settings (Gramann et al., 2011; Kerick & McDowell, 2009; Lance, Vettel, Paul, & Oie, 2011; McDowell & Ries, 2013; Merino et al., 2013; Oie & McDowell, 2011). Our stimuli were intentionally connected to the 3D environment used in previous research (Lance et al., 2011; Marathe, Ries, & McDowell, 2013; Vettel et al., 2012) to facilitate a translational pathway from our 2D recognition memory paradigm to planned 3D experiments with increased task complexity. Understanding behavior outside the laboratory will enable development of novel technologies that can monitor and/or improve performance (Kellihan et al., 2013; Lance, Kerick, Ries, Oie, & McDowell, 2012; McDowell et al., 2013), realizing the pragmatic potential of neuroscience in naturalistic settings (Lightfoot, Bachrach, Abrams, Kielman, & Weiss, 2009).

**Acknowledgements**

The authors would like to acknowledge insightful science discussions with colleagues at the Translational Neuroscience Branch at the U.S. Army Research Laboratory, DCS Corporation, and Institute for Collaborative Biotechnologies at University of California, Santa Barbara. This research was sponsored by the U.S. Army Research Laboratory and U.S. Army Research Office, including support by the Institute for Collaborative Biotechnologies, under Cooperative Agreement Numbers W911NF-12-2-0019, W911NF-10-2-0022, and W911NF-09-D-0001. JK was supported by an appointment to the U.S. Army Research Laboratory Postdoctoral Fellowship Program administered by the Oak Ridge Associated Universities through a contract with the U.S. Army Research Laboratory. The views and conclusions contained in this document are those of the authors and should not be interpreted as representing the official policies, either expressed or implied, of the Army Research Laboratory or the U.S. Government. The U.S. Government is authorized to reproduce and distribute reprints for Government purposes notwithstanding any copyright notation herein.

Figure 1

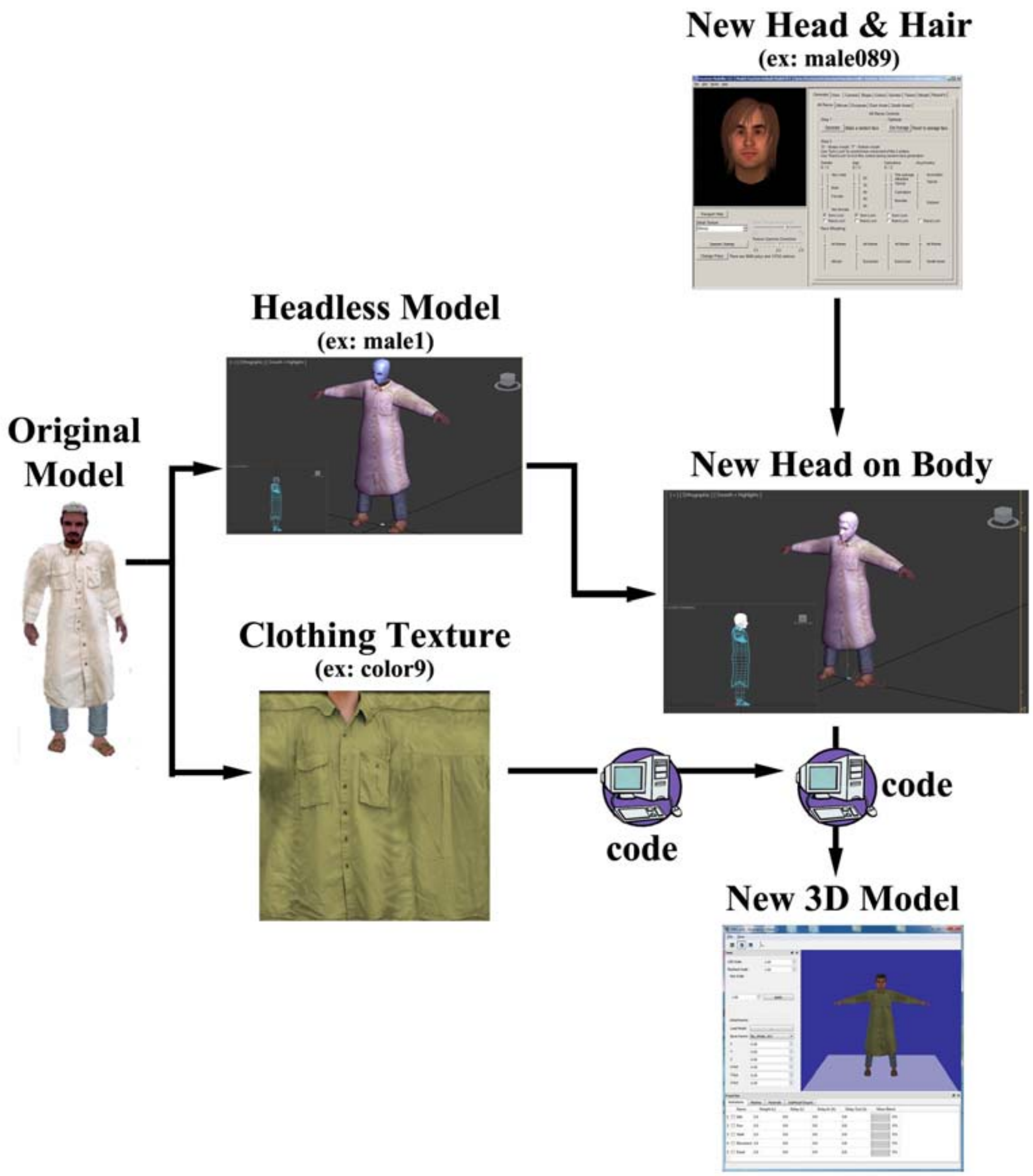

Figure 1. Overview of steps to generate 3D human models. The head model was removed from the purchased human avatar using 3ds Max, and the clothing texture was recolored in GIMP. Concurrently, a new head model and associated hair style was generated in FaceGen and exported in 3ds format. The new head was attached to the avatar skeleton in 3ds Max. The final 3D human entity is a Cal3D model, and it can be used in tools that are built on the Delta3D library, including the Animation Viewer depicted here.

Figure 2

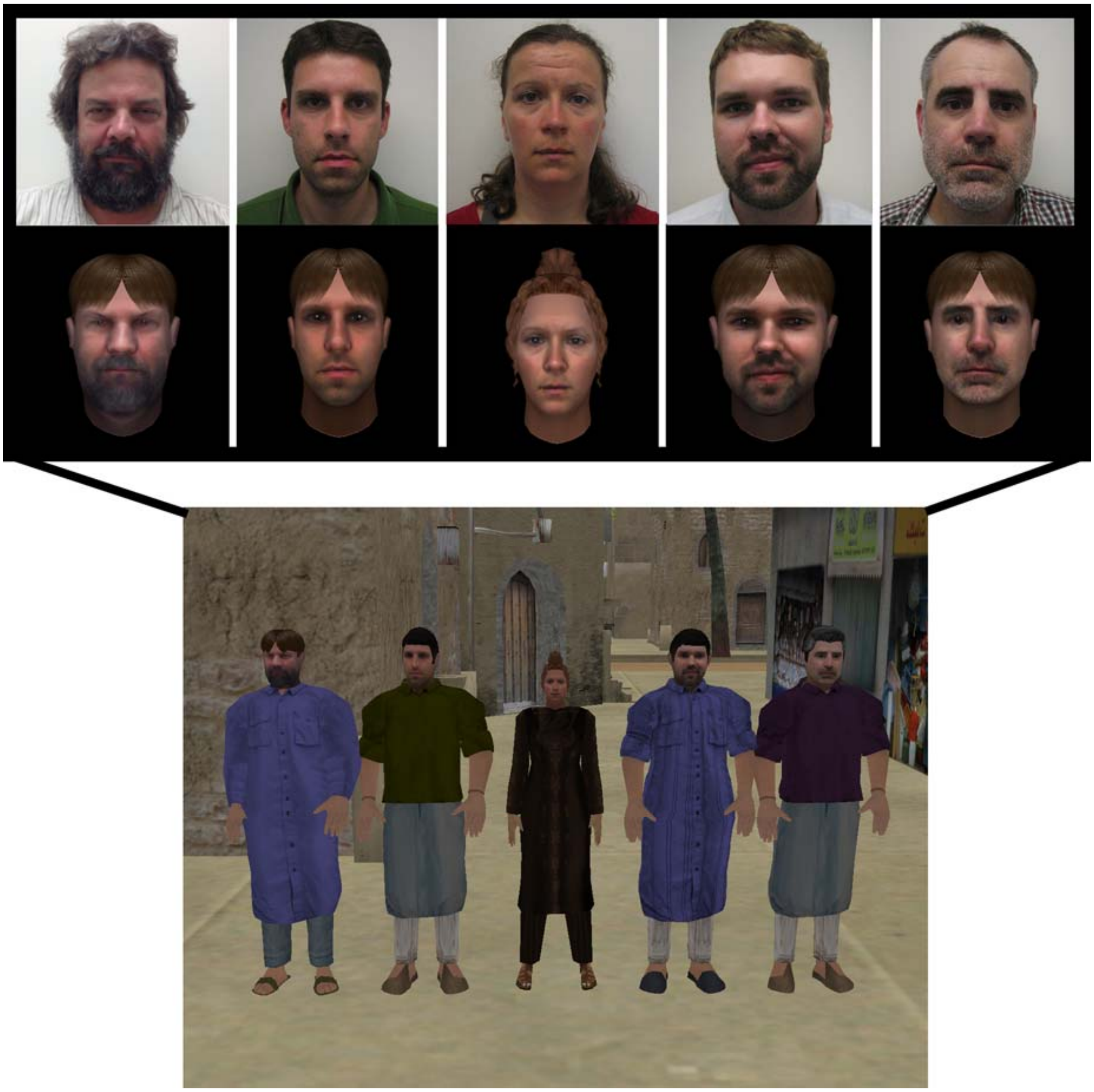

Figure 2 (Top) Front profile photos were used in conjunction with side profile pictures in FaceGen's PhotoFit procedure to create 3D models from personally familiar faces. (Bottom) The five FaceGen models created from photographs are shown here attached to 3D models and embedded in a simulated desert metro 3D environment.

Figure 3

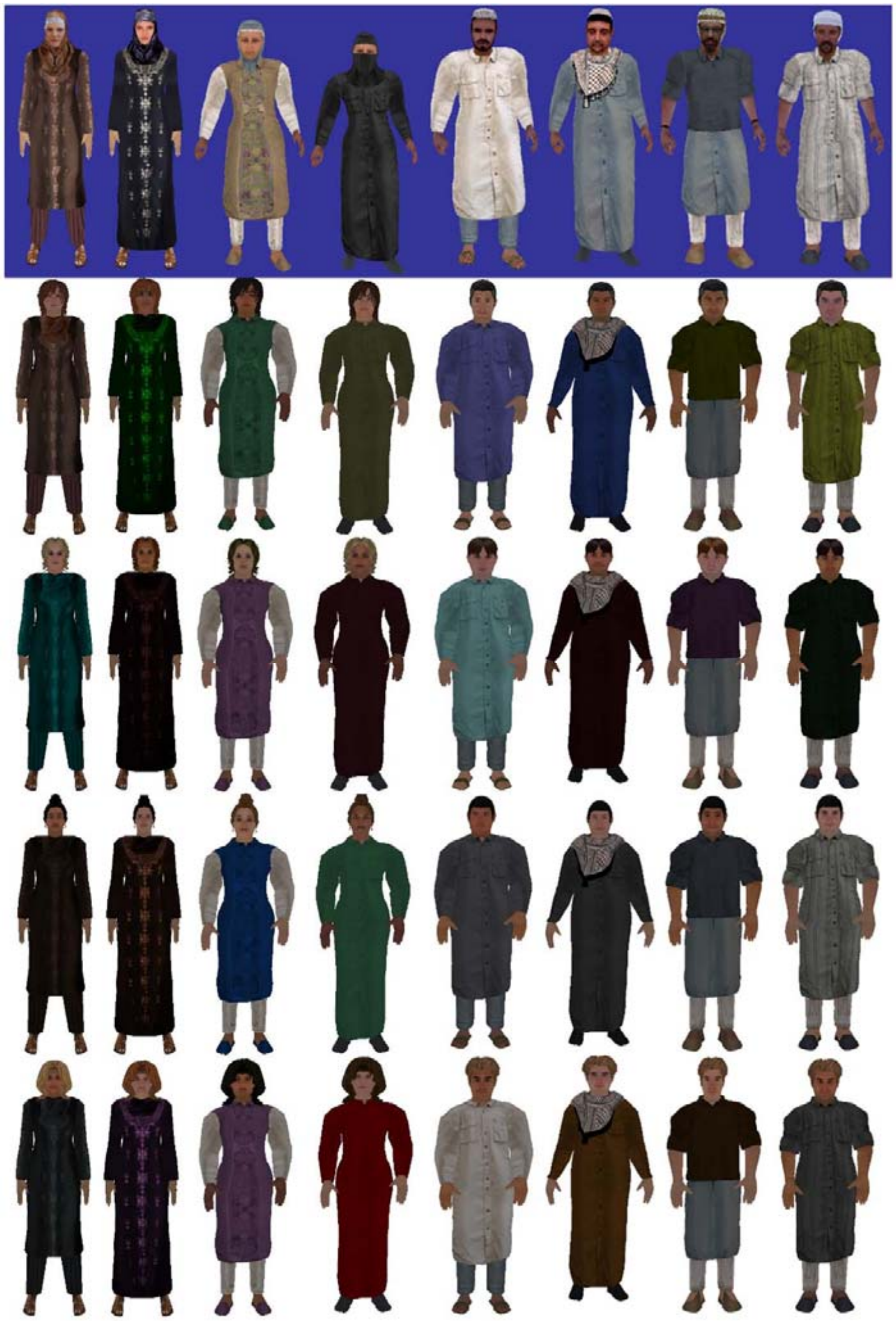

Figure 3. The set of purchased models (top row) and samples of models created for the current set. Each column shows the variability for the model in the set, including a sampling of the clothing colors. Each row represents one of the four female and four male hairstyles.

Figure 4

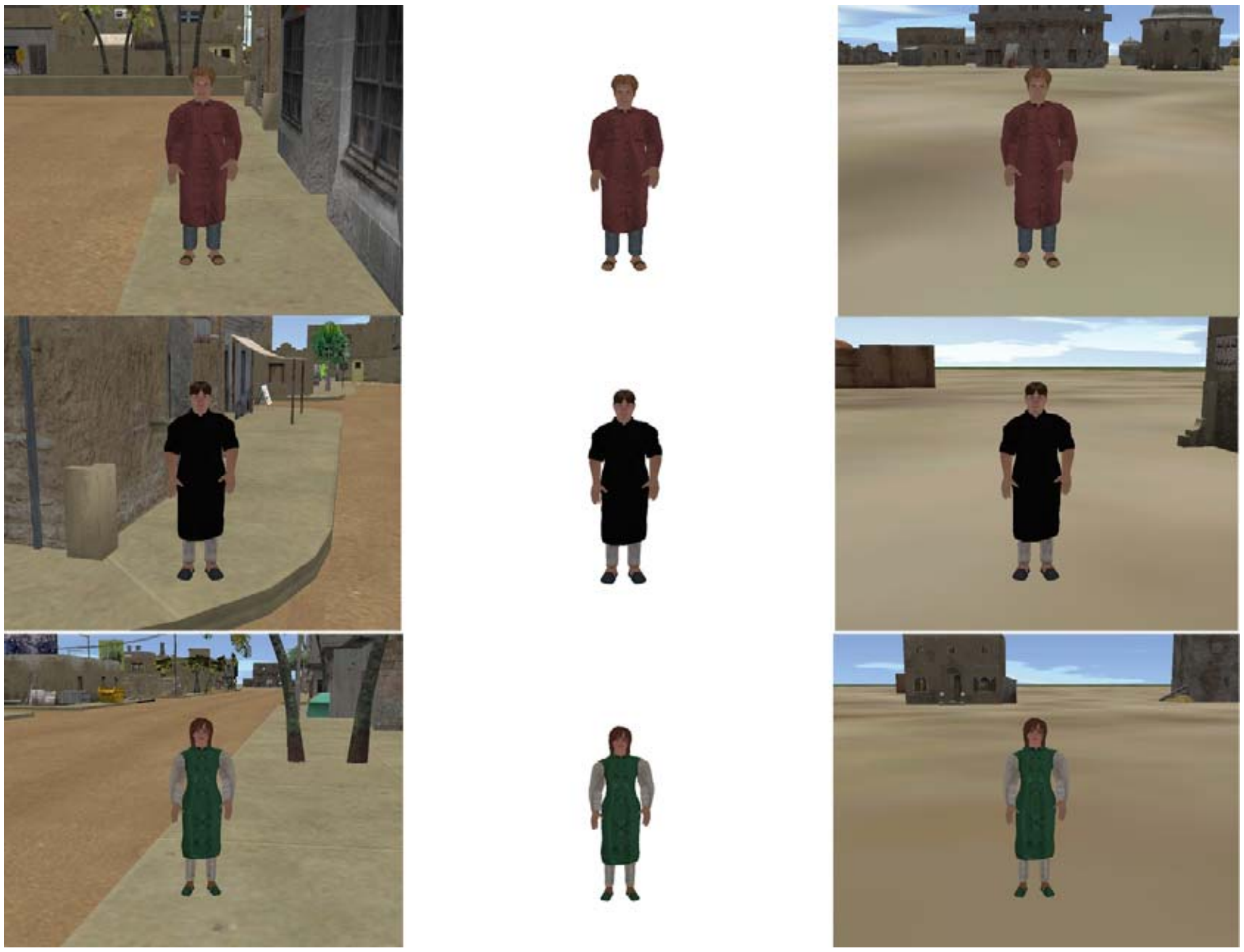

Figure 4. Example models shown in the three background environments: three of eight possible city locations in left column, the white background in the center column, and three of eight possible city outskirt locations in the right column. These 2D images were used in a recognition memory experiment.

Figure 5

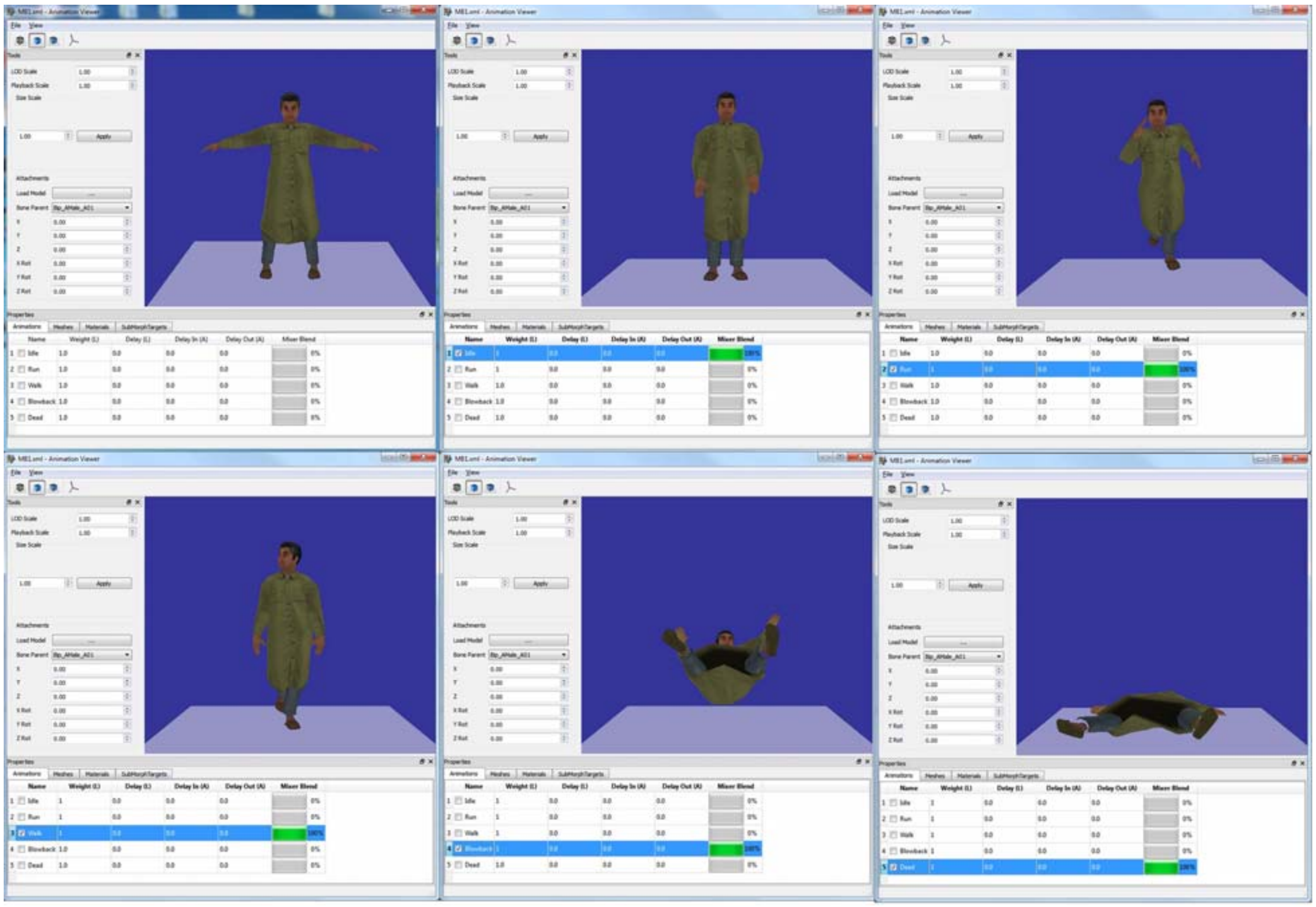

Figure 5. One of the 3D models is shown in Delta3D's Animation Viewer software to highlight the animations available in simulated environments for the new Cal3D models: default posture (row 1, left), standing idle (row 1, center), running (row 1, right), walking (row 2, left), getting blownback (row 2, center), and in a dead position (row 2, right). Future efforts could incorporate customized gesture and action animations.